\documentclass[pdflatex,sn-standardnature]{sn-jnl}
\jyear{2025}
\usepackage{setspace}
\usepackage{lineno}

\usepackage{array,multirow}
\usepackage[normalem]{ulem}
\usepackage{caption}
\usepackage{subcaption}
\usepackage{pifont}

\usepackage{booktabs}
\usepackage{verbatim} 
\usepackage{cleveref}
\usepackage{hyperref}
\usepackage{float}
\usepackage{xfrac}
\usepackage{tabularx}
\usepackage{enumerate}
\usepackage{graphicx}
\usepackage{multicol}
\usepackage{multirow}

\begin{document}

\title[]{\centering Explainability matters: \\ The effect of liability rules on the healthcare sector} 

\author[1]{\fnm{Jiawen} \sur{Wei}}\email{jiawenw@u.nus.edu}

\author[2]{\fnm{Elena} \sur{Verona}}\email{elena.verona@phd.unipi.it}

\author*[3]{\fnm{Andrea} \sur{Bertolini}}\email{andrea.bertolini@santannapisa.it}

\author*[1,4]{\fnm{Gianmarco} \sur{Mengaldo}}\email{mpegim@nus.edu.sg}

\affil[1]{\orgname{Department of Mechanical Engineering, National University of Singapore},\\ \country{9 Engineering Drive 1, Singapore, 117575}}

\affil[2]{\orgname{Dipartimento di Giurisprudenza, Università di Pisa}, \country{Italy}}

\affil[3]{\orgname{Dirpolis Institute, Scuola Superiore Sant'Anna}, \country{Italy}}
    
\affil[4]{\orgname{Department of Mathematics (courtesy), National University of Singapore},\\ \country{10 Lower Kent Ridge Road, Singapore, 119076}}

\abstract{
Explainability, the capability of an artificial intelligence system (AIS) to explain its outcomes in a manner that is comprehensible to human beings at an acceptable level, has been deemed essential for critical sectors, such as healthcare. 
Is it really the case? 
In this perspective, we consider two extreme cases, ``Oracle'' (without explainability) versus ``AI Colleague'' (with explainability) for a thorough analysis.
We discuss how the level of automation and explainability of AIS can affect the determination of liability among the medical practitioner/facility and manufacturer of AIS. 
We argue that explainability plays a crucial role in setting a responsibility framework in healthcare, from a legal standpoint, to shape the behavior of all involved parties and mitigate the risk of potential defensive medicine practices. 
} 


\maketitle


\section{Introduction}


Artificial intelligence (AI) is being increasingly used across various sectors, including some, where decisions may incur liabilities and be subject to legal actions.
One such sector is healthcare. 
Here, decisions made by medical doctors may prove consequential damages to the well-being of a patient. 
Indeed, patients who suffer harm as a consequence of medical treatment might seek compensation, either under tort or contract law. 
The application of AI in the healthcare sector raises the question of \textit{how to determine the liability among medical practitioners, hospitals and artificial intelligence systems (AIS)}.

AI can influence medical liability in multiple ways. 
This work is particularly concerned with the increase in automation that AI provides, and the subsequent collaboration of humans and machines.  
The interplay may result in liability swinging back and forth between the manufacturer of AIS on the one hand, and the medical practitioner and/or facility on the other. 
Physiologically, this very much depends on the level of automation of AIS, its role in the diagnostic/treatment processes, as well as its explainability. 
At the same time, this is also – pathologically – affected by the functioning and failures of the legal system, due to the incentive structures it creates, intentionally or not.

Before moving on to considering how the different aspects of automation, function and degree of explainability affect the application of liability rules and thence determine the incentives that shape the parties’ behaviour, including potential defensive medicine practices, it is necessary to briefly clarify the second dimension just recalled, namely the impact of the legal system. 
The latter is best understood if we consider the following: absent any further specification,
a harmful event such as the failure to correctly treat the patient and/or diagnose a condition might be deemed causally determined either by a choice, decision and/or performance of the practitioner, or by a given output and/or mal- or under-performance of AIS.
While legal criteria might cause different subjects to be held responsible – including, for instance, the medical structure responsible for selecting and maintaining all machinery and applications – the event was materially caused by either one between the practitioner and AIS.
However, even in such cases where the outcome is to be traced back to the malfunctioning of AIS, and therefore the liability should rest with the manufacturer, the complexity of demonstrating constitutive elements of the claim (such as defectiveness or the causal nexus) might induce the victim to seek compensation from the practitioner – or from the medical structure – instead\footnote{One of the most relevant problems whenever advanced technologies are concerned is the distribution of responsibility among potential tortfeasors involved in performing a given action or task while using a partially autonomous device. 
In the case of an increasingly autonomous vehicle, a given accident might be the consequence of (i) the autonomous driving system failing to avoid an obstacle or drive safely, (ii) of the human driver failing to supervise the system, or erroneously deciding to activate it when it would have been safer not to do so, or (iii) of the smart infrastructure failing to provide information about the position of other vehicles, or (iv) of the internet service provider ensuring such connection. 
The complexity in attributing liability to one of parties depends on both practical aspects, such as the difficulty in determining the exact causal series of events that lead to the accident, and the contribution of each individual party towards the result, and the difficulty in applying existing legal norms, demonstrating the constituting elements of liability such as the defectiveness (of the product) or the fault (of the human agent). 
For a detailed discussion, please refer to~\cite{bertolini2021grounding}.}.
Said otherwise, legal norms are not neutral and do provide incentives that might align or not with the position of the claimant.
Therefore, if establishing that the AIS used in the diagnostic process failed to perform due to its inherent defects proves excessively challenging, the party could seek compensation from the medical practitioner for reasons of simplicity~\cite{bertolini2021grounding}.

As per the non-legal aspects recalled, some fundamental considerations may be drawn.
All AIS in healthcare are today designed as tools to assist  practitioners. 
Their autonomy is thence limited and does not allow for a final decision to be made by the machine, irrespective of human supervision and control. 
Such form of ``weak autonomy''~\cite{gutman2012action} certainly reduces the scope of liability for the designers and manufacturers of AIS, in as much as there is always a human user in control. 
To the contrary, an increase in automation, if it progressively carved out room for independent judgments to be reached without human supervision, would expand the domain of manufacturers’ liability. 
From this perspective, it shall be observed how both (i) the way in which AIS is used and the function it serves, and (ii) the degree of transparency and explainability of AIS affect the behaviour of the human user in deciding whether to adhere to the recommendations given by AIS, beyond the formal degree of automation that AIS itself possesses. 
In other words, even if AIS were never conceived to be fully autonomous, the real extent of independence of the human judgment and conclusion greatly varies depending, for instance, on whether the assessment of AIS precedes or follows that of the user, as well as on the degree of explainability reached.

As per the role and function of AIS in the diagnostic process, some empirical research in radiology has shown that greatest human independence in judgment is ensured when AIS serves as a second reader, as opposed to first reader or even filtering system~\cite{regge2013cad, amore2023robotica}. 
When the machine provides a first assessment or filters the images that the practitioner is required to evaluate – for instance excluding clear negatives – the likelihood that the human will deviate from the automated recommendation, or even decide to check and revise the filtering operated by the machine, is more limited. 
In such cases, even if AIS was presented as only partially autonomous, requiring human supervision and control, the result is that of inducing or favouring adherence to the automated output. 

\begin{figure}
    \centering
    \includegraphics[width=\linewidth]{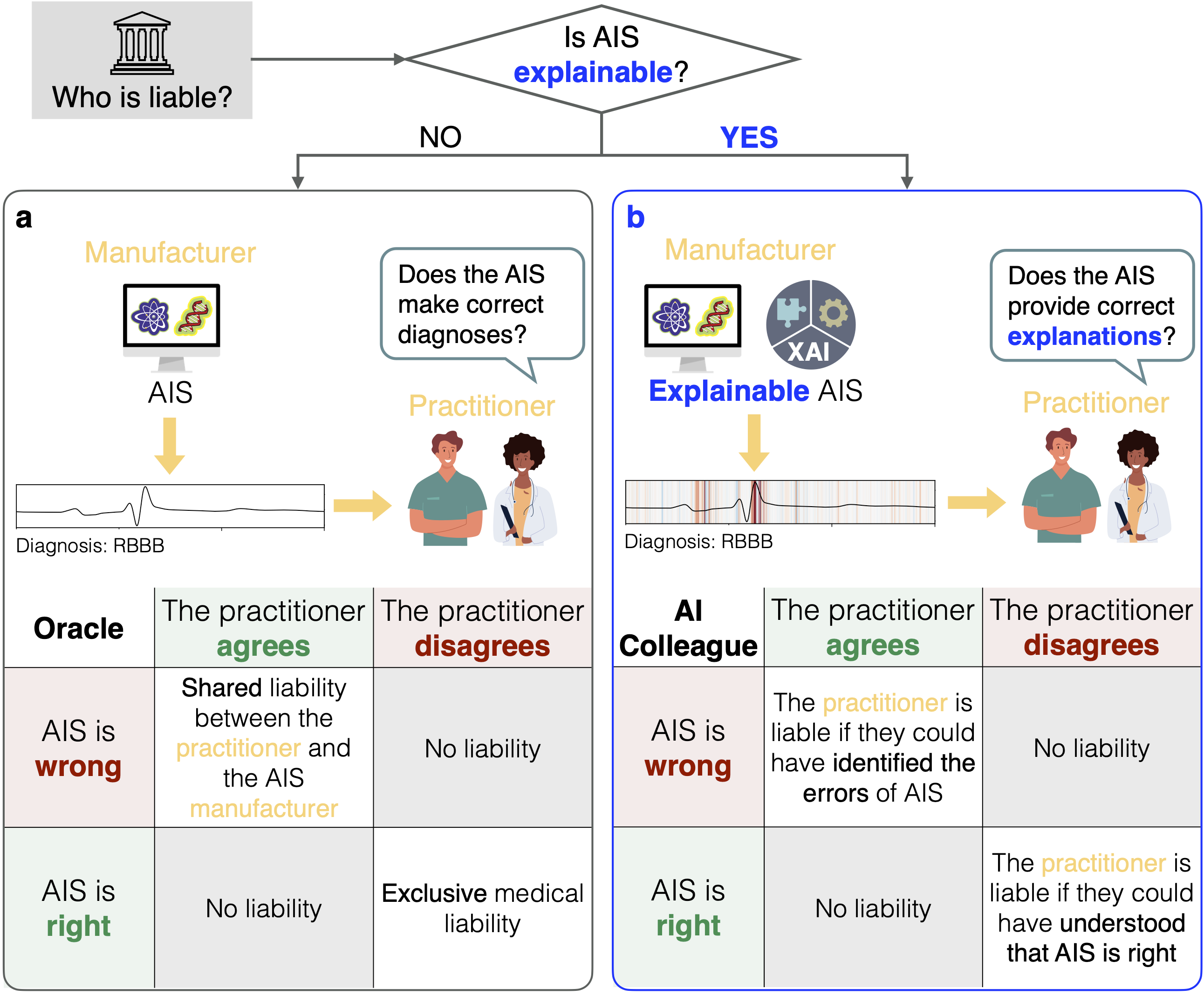}
    \caption{The responsibility framework for artificial intelligence systems (AIS) in healthcare. \textbf{a}, The liability distribution is influenced by the level of automation and explainability of AIS. \textbf{b}, The potential liability for the parties in \textbf{``Oracle''} -- the opaque AIS without any explainability. \textbf{c}, The potential liability for the parties in \textbf{``AI Colleague''} -- the explainable AIS that can explain its conclusions, allowing the practitioner to review outcomes like they would with a colleague's diagnosis.}
    \label{fig:diagram}
\end{figure}

From a purely legal perspective, following the machine's recommendation appears to be a safer option for minimizing one's liability.
This would entail that medical practitioners may strategically suspend their professional judgment – or limit their independence in thinking and reasoning – to the detriment of the patient, to adopt a strategic behaviour primarily addressed at limiting potential responsibility. 
In this context, AI-supported diagnosis could rapidly become the new frontier of defensive medicine (see Subsection~\ref{subsec:dm}).
To better understand such concern, it is necessary to analyze in greater detail the two extreme cases of a completely non-explainable AIS, and an explainable one. 
We call the first ``\textbf{Oracle}'', while the second ``\textbf{AI Colleague}'' (see Fig.~\ref{fig:diagram}).

For the sake of this work, an explainable AIS is either constructed with procedural steps that inherently make it interpretable, or is interpreted a posteriori via suitable methods. 
The former is commonly known as ante-hoc interpretability~\cite{rudin2019stop}, while the latter as post-hoc interpretability~\cite{turbe2023evaluation}. 
They both target the aspect of outcome interpretability, that is, how the AIS reach a certain decision, for instance what data the AIS used, and how. 
The notion of explainability is directly linked to interpretability, whereby human experts (medical doctors in the case of healthcare) can understand why the AIS reached a certain decision by tapping into their domain knowledge. 
Therefore, they can explain AIS decisions~\cite{mengaldo2024explain}.
We note that we make a clear distinction here between interpretability and explainability. 
The first is merely a way of producing a machine view that human practitioners can then use to achieve the second -- i.e., explaining AIS decisions~\cite{mengaldo2024explain}. 

We argue that explainability, and not just mere interpretability, from a legal perspective, is the key for setting a responsibility framework.
Before outlining the reasoning behind this argument, we introduce some concepts of liability in the healthcare sector.

\section{Liability in the healthcare sector}

\subsection{The different liability rules: an overview}
Liability in healthcare is usually related to harm inflicted on patients due to medical treatments.
Patients can seek compensation under two legal constructs: tort or contract law.
Practitioners typically respond under fault-based tort liability requiring proof of negligence, whereas hospitals usually face contractual liability.
Solutions differ from country to country~\cite{zerbo2020guidelines, koch2011medical, stauch20152013}, where policy decisions aim to find a trade-off between patient compensation and prevention of defensive medicine~\cite{bester2020defensive}.

Contractual liability typically presumes fault, placing the burden on practitioners to prove diligence.
Additionally, statutes of limitations are usually longer for contractual claims\footnote{Under Italian law, limitations are 5 years for tort (Art. 2947 Italian Civil Code) and 10 years for contracts (Art. 2946 Italian Civil Code); in France, 10 years from injury consolidation (Art. L1142-28 Code de la santé publique); in Germany, generally 3 years (§195 BGB) from the claimant's knowledge of injury and responsible party, or from when they reasonably should have known (§199 BGB).}, thus facilitating malpractice cases, where harm may manifest late, due to illness latency and assessment difficulties.

Tort liability, instead, applies in the absence of a contractual agreement, holding the party responsible who caused damage, traditionally due to their negligence\footnote{In the US, medical malpractice is a separate category of tort: the practitioner’s liability arises from negligence, i.e., the failure to exercise reasonable care, if the breach of the standard of care was the cause of the patient’s injury. The existence of a contract between the practitioner and the patient holds limited importance, as the doctor cannot contractually modify or eliminate this duty, although many states impose a cap on non-economic or total damages that limits medical liability~\cite{hyman2012medical}.}. 
Negligence is defined as the failure to exercise reasonable care in specific circumstances. 
To define negligent behaviour in the medical profession, reference is made to medical guidelines and best practices: the model of due care is usually a conscientious average doctor in the actual situation~\cite{koch2011medical}. 
While compliance with medical guidelines typically does not completely eliminate the liability of the professional, it certainly provides a stronger starting position for the doctor’s defence.


This aspect tightly intertwines with another element a claimant needs to demonstrate, specifically causal nexus, that entails showing that the damage results from the practitioner’s conduct. 
Causation is particularly difficult to prove, as it often involves distinguishing the effects of malpractice, the usual risks associated with medical treatments, and the patient's preexisting conditions\footnote{To reduce the hurdle the plaintiff has to sustain, each jurisdiction adapts its rules of evidence: for instance, in France and in Italy causation can be proved by presumptions as long as they are serious, precise and concordant; in Germany, Section §630 h establishes that causation is presumed when the practitioner has committed gross negligence (\textit{grobe Behandlungsfehler}) that is likely to cause an injury.}. 
Often, it is uncertain whether a different action by the practitioner would have improved the patients' condition or increased the chance of curing their illness.
That is why, many countries, such as France and Italy, along with many US States, allow compensation for a mere ``loss of chance'', defined as the reduction in the probability of a patient's recovery or improvement due to the practitioner’s faulty performance~\cite{oliphant2013medical}.



From a strategic game-theoretic standpoint, compliance with guidelines and practices will prove the dominating strategy to minimize liability risks, especially in uncertain situations, potentially leading to defensive medicine practices.
In this context, similarities may be drawn between guidelines and AIS used in diagnosis, in particular when less explainable solutions are considered (see Section~\ref{sec:AIS} below).

Instead, practitioners are generally not responsible for AIS failures that can be traced back to the training and design.
In such cases, product liability holds manufacturers liable for user damages, regardless of negligence.
The rationale is that those who profit from distributing a product should also internalize any negative economic consequences from its use, provided the product is considered ``defective''~\cite{Owen_2008}.

While certifying the product as safe is a precondition for its legal commercialization, it does not exclude that the possibility that the product may be defective, or the subsequent duty to compensate for any resulting damages.
In fact, defectiveness may be defined in different ways. 
While US legal doctrines traditionally distinguish different kinds of defects (manufacturing, design and information) (see~\cite{Owen_2008} and Restatement Third of Torts: Product Liability, 1998), Europe refers it to the lack of safety which a consumer is entitled to expect, considering all relevant circumstances, including the product's presentation and characteristics, its reasonably foreseeable use, and the time it was placed on the market or out of the manufacturer's control (art. 7 Product Liability Directive 2024/2853). 
Thus intended, defectiveness fails to capture inadequate performance that does not compromise safety.
It could thence be disputed whether an algorithm used for diagnosis that fails to identify a given lesion is defective, for its inadequacy pertains the quality of its performance rather than a lack of safety per se~\cite{machnikowski2016european}.

The previous regulatory framework for product liability demonstrated limited effectiveness in ensuring victim compensation. 
According to a study conducted in 2018 \cite{eu_commission_2018}, only 547 cases were litigated across 25 European Member States (MS) from 2000 to 2016, with many often solved by resorting to general tort and contract law norms.
The opposite is true in the US, where product liability claims are frequent and ensure substantial compensation for victims~\cite{machnikowski2016european}.
One reason for the limited application of product liability in Europe is traditionally attribute to the high costs of litigation, which grow alongside technological complexity~\cite{eu_commission_2018}. 
Consequently, in cases where liability may be distributed among multiple uncertain responsible parties, claimants will likely seek compensation from others rather than the manufacturer, eventually the medical professional or hospital. 


\subsection{The issue of defensive medicine}
\label{subsec:dm}

Liability rules have proven to profoundly influence medical practitioners' behaviour in many aspects.
Fear of – often excessive – litigation may sometimes discourage practitioners from undertaking medical specializations (e.g., emergency, general, orthopedics and neurosurgery, obstetrics/gynecology, radiology), and also influence their performance and professional decisions regarding treatment~\cite{studdert2005defensive, hiyama2006defensive}.

This phenomenon, known as ``defensive medicine'', entails a deviation from the usual behaviour or established good practices, for the sole purpose of minimizing the risk of lawsuits and complaints by patients and their families~\cite{ortashi2013practice}.
The behaviour may involve ordering unnecessary tests, undertaking practices that are not instrumental to treatment, or radically rejecting treatment for high-risk patients or complex clinical situations~\cite{bester2020defensive, ortashi2013practice}.
This shall not be confused with careful practice, where the doctor may request additional tests to understand the patient's condition better in a complex case, as the latter ultimately benefits the patient~\cite{de1998act}.

Typically, defensive medicine increases healthcare costs and might lead to the overuse of unnecessary medications and procedures, eventually limiting access to essential and scarce medical care~\cite{hermer2010defensive}.
The phenomenon is widespread, albeit varying among medical specialties and clinical settings, and is a common pattern across most countries, including the US~\cite{studdert2005defensive}, UK~\cite{ortashi2013practice}, Japan~\cite{hiyama2006defensive}, as well as European countries~\cite{garattini2020defensive}.


Given the clear and strong positive correlation between medical malpractice claims and defensive medicine, legal reform of existing tort law systems is traditionally considered an essential solution.
However, while the positive effect of specific normative interventions is sometimes questionable – and a multidisciplinary approach is definitely needed~\cite{antoci2022curing} –, it is certain that any aspect expanding the practitioner’s exposure further exacerbates the phenomenon.

Next, we will further demonstrate that using opaque AIS may increase practitioners' defensive behaviour, as they are incentivized to always comply with AIS solutions regardless of the accuracy (see Subsection~\ref{subsec:oracle}). In contrast, implementing reliable explanations of how AIS operates restores more control to practitioners, enabling them to leverage their expertise for a thorough assessment of AIS outcomes (see Subsection~\ref{subsec:colleague}).

\section{How AIS explainability can affect the practitioner behaviour}
\label{sec:AIS}

We distinguish two cases below: (i) the ``\textbf{Oracle}'', where the AIS is a black-box without providing any insights into decisions, and (ii) the ``\textbf{AI Colleague}'', which is a fully explainable AIS.
These two cases highlight the importance of explainability.

\subsection{``Oracle'' -- The opaque AIS}
\label{subsec:oracle}

If one AIS acts as an ``Oracle'' -- providing diagnoses or conclusions based on the practitioner's input or directly collected patient data without any explanation -- and the practitioner decides whether to follow its advice, potential liabilities are summarized in Fig.~\ref{fig:diagram} panel \textbf{b}.

In this case, the medical practitioner cannot distinguish when the system is right or wrong beyond their personal expert assessment of the patient. 
Even with low AIS accuracy – i.e., probability of the AIS being right – if the practitioner were to make decisions after receiving the Oracle's output, they would benefit more by strategically aligning with the AIS, placing themselves in the first column.
If the AIS was right, the practitioner would not be liable; if wrong, liability could be potentially shared between the practitioner and the manufacturer, or even shifted entirely to the latter. 
More precisely, the following \textit{considerations} could be made:
\begin{enumerate}[(a)]
    \item \label{considerations(a)} Medical malpractice is established when the practitioner is deemed negligent. 
    Negligence is defined as an unjustifiable deviation from a standard of conduct that is deemed demandable in given circumstances. 
    In this context, an AIS reaching the same conclusion as the practitioner could most likely constitute a \textit{prima facie} case to exclude negligence and/or reproachable deviation from the demandable conduct.
    In other words, unless the AIS is obviously wrong, to the point that a diligent practitioner would notice, the practitioner could ground a claim whereby their concordance with the system demonstrates that the conclusion is not due to negligent or unreasonable behaviour.
    \item Moreover, one could argue that the AIS is defective and led the practitioner to be mistaken, and therefore liability should rest solely with its designer/producer/manufacturer. 
    The latter is even more true when it is the AIS itself that acquires the data. 
\end{enumerate}
We can therefore build a game-theoretic model by assigning hypothetical payoffs for simplicity:
\begin{itemize}
    \item \textbf{No Liability} $\rightarrow$ 0 
    \item \textbf{Shared Liability} $\rightarrow$ $-L \leq X \leq 0$
    \item \textbf{Exclusive Medical Liability} $\rightarrow$ $-L$ (full responsibility)
\end{itemize}
Define $p$ as the probability of AIS being correct (accuracy of AIS). 
Since the practitioner does not know if the AIS is correct beforehand, the probabilities for correctness and errors are $p$ and $1-p$, respectively.
If the practitioner agrees with the AIS, the expected payoff ($E_A$) is equal to: 
\begin{equation}
    E_A = p*0+(1-p)*X=X-p*X.
\end{equation}
If the practitioner disagrees with the AIS, the expected payoff ($E_D$) is equal to:
\begin{equation}
    E_D = p*(-L)+(1-p)*0=-p*L.
\end{equation}
We can find the value of $p'$ where the practitioner is indifferent to either strategy by solving $E_A = E_D$:
\begin{equation}
    p=\frac{X}{X-L}
\end{equation}
If $X$ is close to 0, the practitioner will agree with the AIS even at low accuracy ($p'\rightarrow0$).
If $X$ is close to $-L$, agreement with the AIS requires an accuracy exceeding 50\% ($p'\rightarrow0.5$).
Assuming an unfavourable shared liability condition for the practitioner, an accuracy below 50\% would suffice for agreement as a dominant strategy to reduce potential liability.
However, a system with such low accuracy is unlikely to be certified and used\footnote[14]{The International Organization for Standardization (ISO) is developing the ISO/IEC 42000 series of standards to help organizations to develop safe AI systems responsibly and effectively: the most important standard at the moment is the ISO/IEC 42001, published in 2023, that specifies requirements for establishing, implementing, maintaining and continually improving an Artificial Intelligence Management System (AIMS) within the context of an organization (\url{https://www.iso.org/standard/81230.html}).}.

The more accurate the AIS – beyond 50\% – the preferable the strategy to agree every time. 
Moreover, it shall be further noted that, the efficacy of the reasoning summarized under \textit{consideration} (\ref{considerations(a)}) above, and therefore the likelihood of escaping liability when the practitioner nonetheless agrees with a wrong AIS, increases as the accuracy.
Higher AIS accuracy further increases the payoff $X$, thence increasing the possibility of shifting liability to the manufacturer. 
Consequently, the dominant strategy is even more that of complying with AIS conclusions.

However, if we set aside the practitioner's perspective and their interest in minimizing potential liability, we might notice that an accuracy around 50\% should not induce the practitioner to rely on the system rather their own expertise and knowledge. 
It would be in the patient’s interest if the doctor exerts entire discretion and adhere solely to professional standards. 
Said otherwise, a strategic agreement with the AIS, indeed, can be viewed as a form of defensive medicine.

\subsection{``AI Colleague'' -- The explainable AIS}
\label{subsec:colleague}

If we instead assume a completely explainable AIS that can justify its conclusions as precisely as a doctor would, allowing the practitioner to review and assess outcomes like they would with a colleague's diagnosis, the functioning of liability rules would vary.
The more explainable the AIS is, the more practitioners become liable for diagnostic errors, positioning AIS merely as a tool they use. 
Specifically, explainability would impact the level of care expected from the practitioner when they are not fully convinced by the explanation.
We call such an explainable AIS an ``AI Colleague''.

For the sake of clarity, we concentrate on the liability of the practitioner. 
Unlike the case of the ``Oracle'' above, there is no shared liability hypothesis in ``AI Colleague''.
The practitioner holds full responsibility (payoff $-L$) for any mistakes in decisions, and incurs no liability if they disagree with a wrong explainable AIS. 
Note that mere disagreement may not suffice to exclude the liability in practice; nonetheless, this is a negligible aspect for now considering the focus of this analysis.





The practitioner could be held exclusively liable if they agree with a wrong AIS, as the explanation provided should clarify the reasoning behind the solution.
Here, the doctor is put in the position to assess clinical reliability of the machine's diagnosis based on existing guidelines and good clinical practices, just as they would for a colleague's recommendations. 
Symmetrically, if the AIS is right but the doctor disagrees, the practitioner would be held liable for the wrong diagnosis, given that correct reasoning provided by the explainable AIS should have induced them to agree. 
Therefore, the standard of care expected from the doctor entails determining if they could have identified the errors or understood the correctness of the explainable AIS, another colleague under similar circumstances.
The potential liabilities in ``AI Colleague'' are summarized in Fig.~\ref{fig:diagram} panel \textbf{c}.

In ``AI Colleague'', an AIS with accuracy $p \geq 0.5$ suggests the practitioner is better off agreeing with the system, as we stated in Subsection~\ref{subsec:oracle}, it is also true that liability directly depends on the practitioner's ability to detect the errors of AIS.
This also allows practitioners to reduce or potentially radically exclude liability by identifying AIS errors and disagreeing correctly, rather than relying on identical accuracy $p \geq 0.5$.
The ability to identify errors – said otherwise, the practitioner's accuracy – can be denoted as $d_m (0 \leq d_m \leq 1)$.

We can further clarify that the practitioner’s ability to detect mistakes depends on accuracy $p$ and explainability $E (0 < E \leq 1)$ of the AIS. 
If the system is perfectly explainable, $E = 1$.
It is reasonable to assume that a more accurate AIS, coupled with a clearer and more thorough explanation of its reasoning, enhances the practitioner's ability to identify errors.
The individual practitioner's purely subjective propensity to detect errors can be represented by a coefficient $k (0 \leq k \leq 1)$, reflecting their diligence as a percentage of that of a perfectly diligent practitioner. 
The more diligent, apt, and attentive the individual practitioner, the closer the coefficient $k$ will approach 1.
We may therefore define $d_m$, as: 
\begin{equation}
    d_m = f(p, E, k)=k*p*E
    \label{eq:dm}
\end{equation}

There also exists a probability that the practitioner mistakenly judges the AIS is wrong, given a right AIS with completely correct explanation, which can be denoted as $w (0 \leq w \leq 1)$.
Similarly, the variable $w$ is influenced by diligence coefficient $k$ of the practitioner, as well as the accuracy $p$ and explainability $E$ of the AIS.
We may therefore define $w$, as:
\begin{equation}
    w=g(p,E,k)=(1-k)*(1-p*E)
    \label{eq:w}
\end{equation}

\begin{figure}
    \centering
    \includegraphics[width=0.6\linewidth]{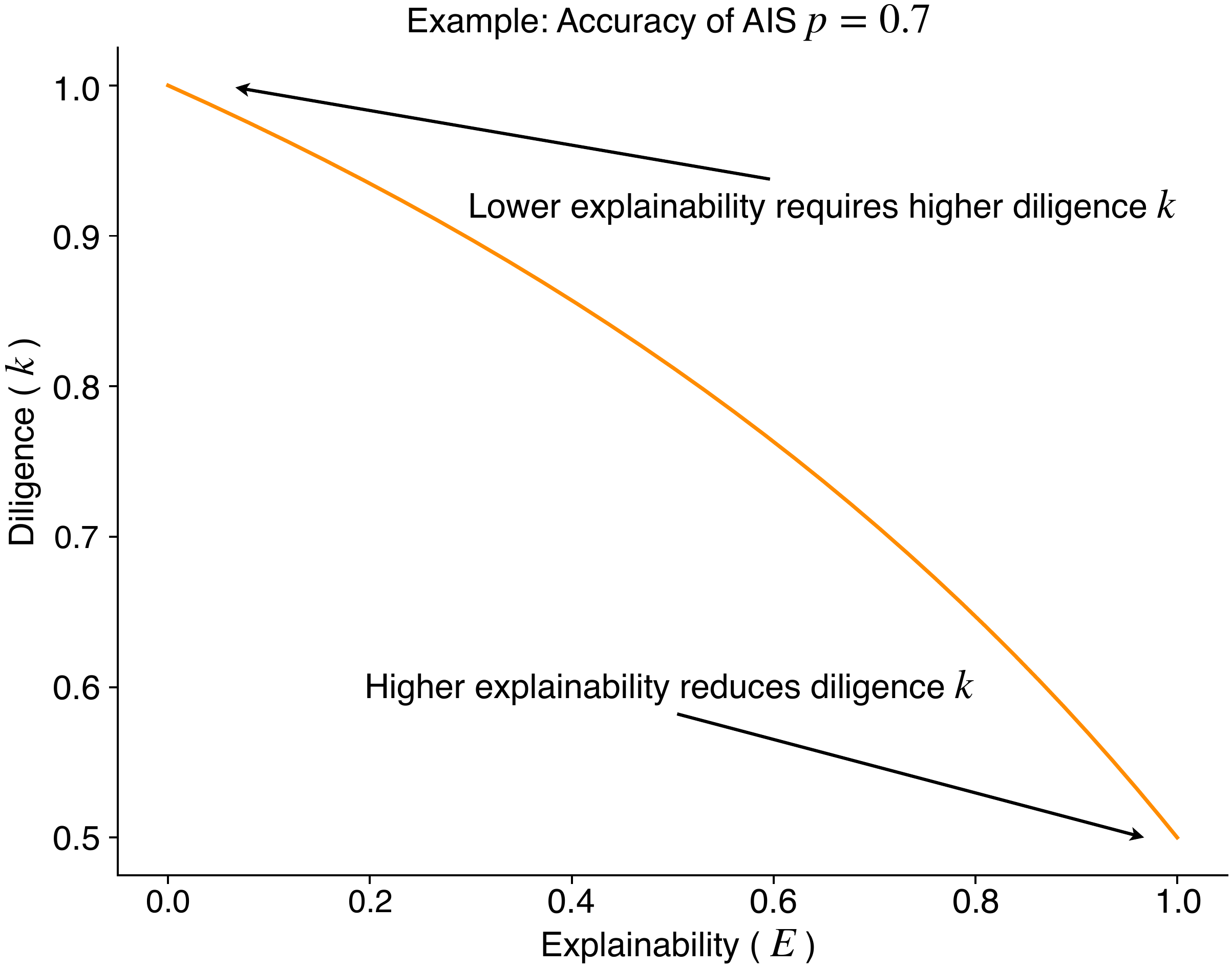}
    \caption{The relationship between the diligence ($k$) of the practitioner and the explainability ($E$) of AIS, given an accuracy of 70\% ($p=0.7$) for example. AIS with lower explainability requires higher diligence from practitioners, while higher explainability reduces the diligence required.}
    \label{fig:kE}
\end{figure}

Next, we can develop a game-theoretic model for ``AI Colleague'', where the practitioner chooses between two strategies: 1. Systematic Agreement (systematically agree with the AIS), or 2. Independent Assessment (independently assess the AIS outcomes prior to agreeing or disagreeing).
The payoff structure of the model is outlined in Table~\ref{tab:payoff}.
%
\begin{table}[ht]
    \centering
    \caption{The payoff structure of ``AI Colleague''}
    \begin{tabular}{|l|c|c|}
    \hline
        \multirow{2}*{\textbf{Explainable AIS}} & The practitioner agrees  & The practitioner assesses  \\
        & systematically & independently \\
        \hline
        The AIS is wrong & $(1-p)*L$ & $(1-d_m)*(1-p)*L$ \\
        \hline
        The AIS is right & 0 & $w*p*L$ \\
        \hline
    \end{tabular}
    \label{tab:payoff}
\end{table}
%
%
\begin{enumerate}
    \item \textbf{Systematic Agreement} \\
    Liability arises only when the AIS is wrong, namely $(1-p)$, resulting in the expected liability of $E_{SA}=(1-p)*L$.
    \item \textbf{Independent Assessment} \\
    Liability arises when the practitioner (i) fails to identify errors of the AIS, with a corresponding payoff of $(1 - d_m) * (1 - p) * L$, and (ii) misjudges correct AIS outcomes as wrong, with a corresponding payoff of $w * p * L$. 
    By substituting $d_m$ and $w$ from Eq.~(\ref{eq:dm}) and (\ref{eq:w}), respectively, the expected liability can be calculated by $E_{IA}=(1 - k * p * E) * (1 - p) * L + [(1 - k) * (1 - p * E)] * p * L$.
\end{enumerate}
%
%
By solving $E_{SA}=E_{IA}$:
\begin{equation}
    (1 - p) * L = (1 - k * p * E) * (1 - p) * L + [(1 - k) * (1 - p * E)] * p * L
\end{equation}
we can find the threshold value of $k'$ in indifference condition,
\begin{equation}
    k = \frac{1-p*E}{(1-p)*E+1-p*E}
\end{equation}
When AIS is highly explainable ($E \rightarrow 1$), the threshold value of $k$ for independent assessment by the practitioner approaches 0.5.
This means that, for a highly explainable AIS, given its accuracy, even a less diligent, skilled, and attentive practitioner is better off identifying AIS errors and undergoing an independent assessment rather than systematically agreeing.
In this way, AIS serves as a tool to support professional expertise, not an ``Oracle'' to be followed uncritically.
Additionally, we can assume $p$ as fixed since only those AIS that fulfill the required accuracy criteria essential for ensuring adequate safety and performance will get certification from competent authorities. 
Therefore, all AIS used by practitioners meet that requirement by definition. 
We show the relationship between diligence $k$ and explainability $E$ under the AIS with accuracy $p=0.7$ in Fig.~\ref{fig:kE}.

In this context, this analysis could compare the behaviour of more practitioners confronted with the option of employing an identical AIS. 
When $p$ and $E$ are fixed as the AIS is identical, the only factor to be considered is the individual's practitioner performance $k$. 
Thus, the overall purpose of the legal system is to ensure that all medical practitioners meet a level of diligence $k \geq k'$ that satisfies the aforementioned condition.
This is ensured both \textit{ex ante} through educational and professional requirements for medical licensing, and \textit{ex post} through liability sanctions.
Ideally, while some practitioners will inevitably be more highly skilled, the majority should at least meet $k'$, and those who fail should be sanctioned through responsibility.
This would, in turn, ensure that AIS serve as tools, while the practitioner retains primary responsibility.

In ``AI Colleague'', liability of AIS manufacturer is to some extent residual. 
The focus would shift from accuracy to erroneous explanations, meaning that the AIS did not fail in diagnosis itself but in explaining its conclusion.
It entails the extreme case where no explanation is provided, ultimately swinging the responsibility back to a diligent practitioner (displaying minimal diligence), who must recognize the malfunctioning AIS and act accordingly.
It may, however, also entail more subtle cases where the AIS provides an explanation that later proved not to accurately match its conclusion. 
In such cases, a discrepancy would arise between the output of AIS and its accompanying explanation, which might not accurately represent the data and/or steps used by the AIS to reach its conclusion. 
Such circumstances could lead practitioners to make mistakes they otherwise would not, swinging responsibility back to the manufacturer of AIS, who could be held exclusively liable for not offering trustworthy explanations.

Indeed, if the provided explanation does not accurately reflect the reasoning of AIS, two different scenarios may occur where the doctor is led to believe that,
\begin{enumerate}
    \item AIS is right, even when it is not, as the provided explanation creates a facade of correctness that may go unnoticed, even by a diligent practitioner;
    \item AIS is wrong, as the provided explanation appears inaccurate, or refers to irrelevant and/or incorrect aspects under the circumstances, inducing a diligent and competent practitioner to doubt the conclusion reached.
\end{enumerate}
Both such cases need to be distinguished, however, from that were the AIS is wrong in its assessment (e.g., erring in identifying a neoplastic lesion).

The relevant aspect here is not the accuracy of the AIS, but the discrepancy between its conclusion and the incorrect explanation, which a diligent practitioner could not have detected.
Therefore, the AI provider could be considered solely responsible for the harm caused by the misinterpretation of the AI solution.

Instead, as previously clarified, the practitioner's responsibility lies in failing to identify the ``hallucination'' or wrong conclusion provided by AIS when they should have, or in not agreeing when the AIS is right.
Both such hypotheses rely on the AIS being explainable.

\section{The new frontier of defensive medicine}
The liability of the manufacturer in ``AI Colleague'' is radically different from that of the practitioner, much more so than in ``Oracle''.
It is potentially more limited to the specific cases outlined, as the decision-making process and the causal series leading to patient harm are clearly distinguishable and observable.

From the manufacturer's viewpoint, explainability takes precedence over accuracy itself once the latter satisfies the AIS certification requirement. 
This is because, \textit{ceteris paribus}, only explainability helps transfer control and subsequently responsibility to the practitioner, avoiding a strategic behaviour that would turn AI diagnosis into the new frontier of defensive medicine practices.

This also provides a clear indication to policy makers that, when identifying relevant criteria for AIS certification in diagnostics, they should not focus solely nor primarily on overall accuracy but also consider explainability.

\section*{Acknowledgments}
We acknowledge funding from MOE Tier 1 grant 22-4900-A0001-0 `Discipline-Informed Neural Networks for Interpretable Time-Series Discovery'.  

\bibliography{references}

\end{document}